\documentclass[%
 aip,
 amsmath,amssymb,
 reprint,%
]{revtex4-1}

\def\qr{\hat{g}^R}

\def\der{\partial_{\vec{R}}}

\usepackage{graphicx}
\usepackage{dcolumn}
\usepackage{float}
\usepackage{bm}

\usepackage[utf8]{inputenc}
\usepackage[T1]{fontenc}
\usepackage{mathptmx}
\usepackage{etoolbox}

\makeatletter
\def\@email#1#2{%
 \endgroup
 \patchcmd{\titleblock@produce}
  {\frontmatter@RRAPformat}
  {\frontmatter@RRAPformat{\produce@RRAP{*#1\href{mailto:#2}{#2}}}\frontmatter@RRAPformat}
  {}{}
}%
\makeatother
\begin{document}

\preprint{AIP/123-QED}

\title{Probing the topological band structure of diffusive multiterminal Josephson junction devices with conductance measurements}
\author{Venkat Chandrasekhar}
 \email{v-chandrasekhar@northwestern.edu.}
\affiliation{ 
Department of Physics, Northwestern University, 2145 Sheridan Road, Evanston, IL 60208, USA
}%

\date{\today}

\begin{abstract}
The energy of an Andreev bound state in a clean normal metal in contact with two superconductors disperses with the difference $\Delta \phi$ in the superconducting phase between the superconductors in much the same way as the energies of electrons in a one-dimensional crystal disperse with the crystal momentum $k$ of the electrons.  A normal metal with $n$ superconductors maps on to a $n-1$ dimensional crystal, each dimension corresponding to the phase difference $\phi_i$ between a specific pair of superconductors.  The resulting band structure as a function of the phase differences $\{\Delta \phi_i\}$ has been proposed to have a topological nature, with gapped regions characterized by different Chern numbers separated by regions where the gap in the quasiparticle spectrum closes.  A similar complex evolution of the quasiparticle spectrum with $\{\Delta \phi_i\}$ has also been predicted for diffusive normal metals in contact with multiple superconductors.  Here we show that the variation of the density of states at the Fermi energy of such a system can be directly probed by relatively simple conductance measurements, allowing rapid characterization of the energy spectrum.         
\end{abstract}

\maketitle

An electron in a clean normal metal (N) in good contact with a superconductor (S) with an energy $E$ less than the superconducting gap $\Delta$ of the superconductor cannot propagate into the superconductor, but is instead reflected as a hole at the NS interface, in a process called Andreev reflection.\cite{deutscher2,andreev}  Andreev reflection is phase coherent, with the phase of the hole and electron related by $\Delta \varphi = - \arccos(E/\Delta) + \phi_s$, where $\phi_s$ is the macroscopic phase of the superconductor.  If a second superconductor is connected to N, the retroflected hole can be transformed by Andreev reflection at the second NS interface to an electron with a similar phase difference; this reflected electron can interfere with the original electron to form Andreev bound states whose energy disperses with the phase difference $\Delta \phi$ between the two superconductors $E = \pm \Delta \cos(\Delta \phi/2)$.\cite{sauls}  The periodicity of this dispersion with $\Delta \phi$ is similar to the periodicity of the levels of electrons in a one-dimensional crystal with the crystal momentum $k$.  The quasiparticle spectrum in a normal metal in contact with $n$ superconductors is a function of the $n-1$ possible phase differences $\{\Delta \phi_i\}$ between them.  Each $\Delta \phi_i$ corresponds to a different `momentum' axis: thus a normal metal with $n$ superconductors in contact corresponds to a $n-1$ dimensional artifical crystal.  Interest in such multiterminal proximity effect Josephson junction (JJ) devices has increased recently since the prediction of the possibility of realizing topologically non-trivial band structures.\citep{amin,heck,riwar,xie,padurariu,amundsen,vischi,xie2,nowak,arnault,huang}  In particular, it has been predicted that for $n\geq4$, the band structure can have Weyl singularities where the gap in the quasiparticle spectrum closes.\cite{riwar,xie2}

In the case of a diffusive normal metal with multiple superconducting contacts, simulations\citep{amundsen,vischi} have shown that one can also tune the quasiparticle density of states at the Fermi energy $N(0)$, resulting in regions in the phase space defined by $\{\Delta \phi_i\}$ where $N(0)$ is reduced or vanishes separated by points at which $N(0)$ approaches the normal state value $N_0$.  Unlike the clean limit, however, $N(0)$ in the diffusive case also has a spatial dependence, typically varying on a length scale defined by the superconducting coherence length in the normal metal (or Thouless length) $L_T = \sqrt{\hbar D/k_B T}$, where $D$ is the diffusion coefficient of quasiparticles in the normal metal and $T$ the temperature.\cite{pannetier,lesueur}  This has two immediate implications for performing experiments to probe the energy spectrum.  First, the geometry of the device is important, and second, the precise location on the device where the energy spectrum is probed is also important.

The latter point is particularly relevant for measurements that probe the energy spectrum using tunneling, as these are sensitive to the density of states at the tunnel contact.  Transport measurements are also sensitive to the density of states.  For example, the thermal conductance of a proximity-coupled diffusive normal metal is directly affected by gaps in the density of states,\cite{jiang} and phase-coherent oscillations in the thermal conductivity of Andreev interferometers have been reported almost two decades ago.\cite{jiang2}  However, thermal conductance measurements on mesoscopic devices at millikelvin temperatures are notoriously difficult to perform.  Here we show through simulations that by measuring the conductance of a suitably designed multiterminal JJ proximity device one can determine local variations in $N(0)$ with $\{\Delta \phi_i\}$ that would normally require local tunneling measurements.  In particular, one can readily identify points in phase space where the gap in the local density of states closes, as the conductance approaches its normal state value at these points.  Conductance measurements thus provide a simple means to explore the artificial band structure of multiterminal JJ devices.

The results presented here were obtained by numerical solution of the quasiclassical equations of superconductivity in the dirty limit, i.e., the Usadel equation\citep{usadel} and corresponding kinetic equations in the Riccati parametrization,\cite{eschrig} for a one-dimensional network geometry, using the open source code developed by Pauli Virtanen.\cite{virtanen,virtanen2}  Details of this formulation are given in the Appendix.  To illustrate the effect of device geometry on the results, we focus here on 2 specific device geometries with 3 superconducting contacts each, which enables us to easily plot the results as a function of the two independent fluxes.  A larger number of superconducting contacts would give rise to a more complex band structure, but qualitatively similar behavior.

\begin{figure}
\includegraphics[width=8cm]{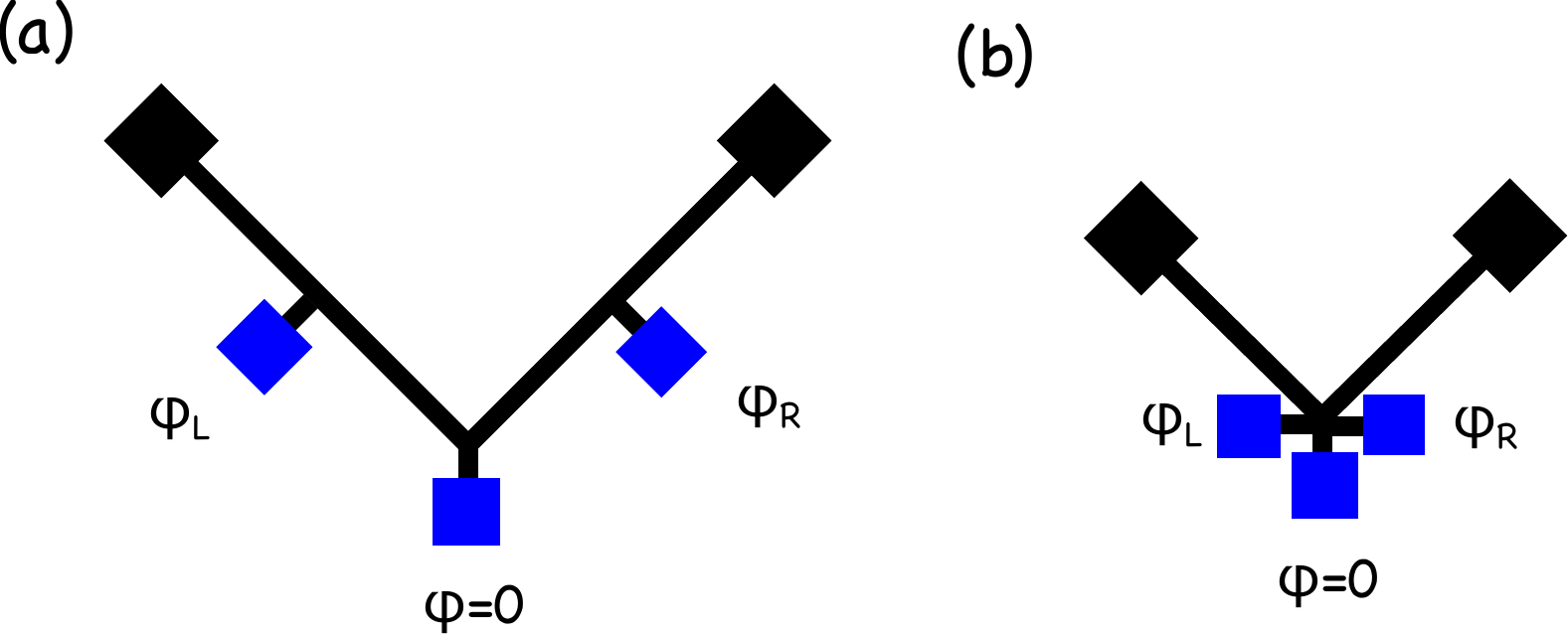}
\caption{Schematic of device geometries for simulations.  Black squares represent normal reservoirs, blue squares represent superconducting reservoirs.  Black lines represent normal wires.  The length of each section of the normal wires between the normal reservoirs is 0.5 $L$, while the short normal section connecting the superconducting reservoirs are of length 0.1 $L$.  Thus the distance between normal reservoirs in device (a) is $2L$, while the distance between the normal reservoirs in device (b) is $L$.  Here $L$ defines the Thouless energy $E_c = \hbar D/L^2$ as discussed in the text.}
\label{fig:fig1}
\end{figure}

Figure \ref{fig:fig1} shows the two device geometries studied here.  Each has three superconducting contacts and two normal contacts.  We arbitrarily set the phase of the middle superconducting contact to be 0, and the phases of the other two superconductors with respect this middle one to be $\phi_L$ and $\phi_R$,  The conductance of the device is measured between the two normal contacts.  The relevant energy scale for the problem is set by the Thouless energy $E_c = \hbar D/L^2$.  Here $L$ is the length of the normal metal $\ldots$ for a device with a single normal metal wire, this is easy to identify, but for a device with normal wires of multiple lengths, the relevant energy scale is not so clear.  Nevertheless, we use this definition of $E_c$, and specify the length of each normal wire in terms of $L$ as noted in the figure caption.  The temperature $T$ and gap $\Delta$ of the superconducting contacts are also given in terms of $E_c$.  For the calculations presented here, we use $T=0.14 E_c$ and $\Delta = 3.18 E_c$, corresponding to experimentally relevant parameters.\cite{noh}  We also assume perfect transparency of the NS interfaces.  A decreased transparency would reduce the variations in the density of states, but would not qualitatively change the results.

\begin{figure}
\center{\includegraphics[width=9cm]{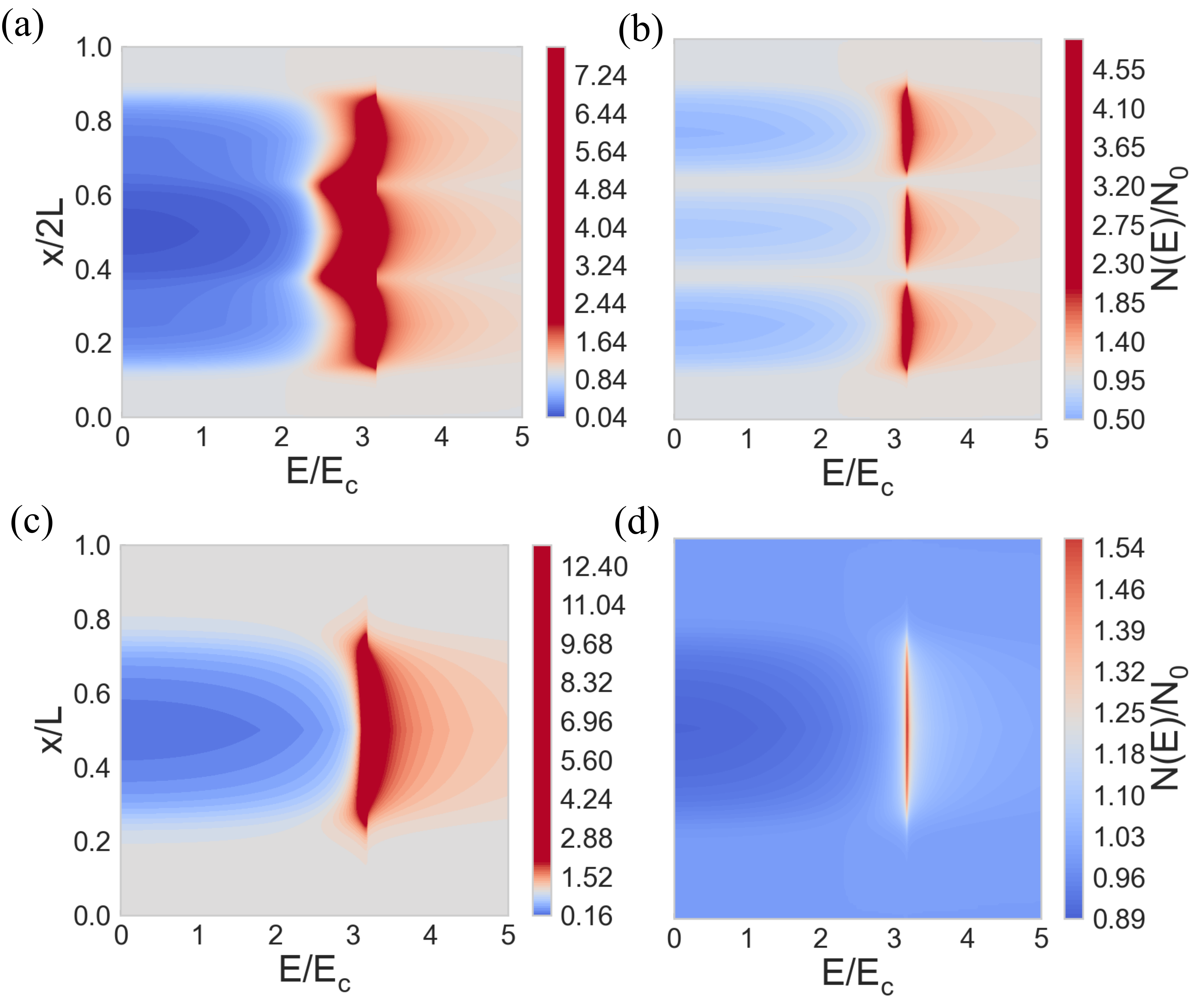}}
\caption{Quasiparticle density of states $N(E)$ as a function of position along the path between the two normal reservoirs.  (a) and (b) correspond to the device geometry of Fig. \ref{fig:fig1}(a) for $\phi_L, \phi_R =0,0$ (a) and $\phi_L, \phi_R =\pi,\pi$ (b) respectively.  (c) and (d) are the corresponding results for the device of Fig. \ref{fig:fig1}(b).}
\label{fig:fig2}
\end{figure}

To  illustrate the effect of the device geometry on the spatial variation of the density of states $N(E)$, Fig. 2 shows $N(E)$ as a function of position along the path between the two normal contacts for the two device geometries of Fig. 1, for two points in the flux phase space:  $\{\phi_L,\phi_R\} = \{0,0\}$ and $\{\phi_L,\phi_R\} = \{\pi,\pi\}$.  The first point corresponds to flux values at which the gap in $N(E)$ should be a maximum and the second to flux values at which one might expect the gap in $N(E)$ to vanish or be a minimum.  The most important fact to note here is that $N(E)$ varies as a function of position for both device geometries, particularly for $\{\phi_L,\phi_R\} = \{0,0\}$, but also more weakly for $\{\phi_L,\phi_R\} = \{\pi,\pi\}$.  The second point to note is that the variation of $N(E)$ with position is more pronounced for the geometry of Fig. 1(a), and that there are points along the length where the gap in the density of states does not close for any $\{\phi_L,\phi_R\}$.  In contrast, for the geometry of Fig. 1(b), the gap in $N(E)$ appears to (almost) close for $\{\phi_L,\phi_R\} = \{\pi,\pi\}$ along the entire length of the wire.  As we shall see below, $\{\phi_L,\phi_R\} = \{\pi,\pi\}$ is near to but not exactly the phase space point at which the gap in $N(E)$ closes completely.  Nonetheless, it is clear from this discussion that appropriate design of the device as well as the exact location that is probed is important in being able to observe a complete closing of the gap as a function of $\{\Delta \phi_i\}$.

\begin{figure}
\includegraphics[width=8cm]{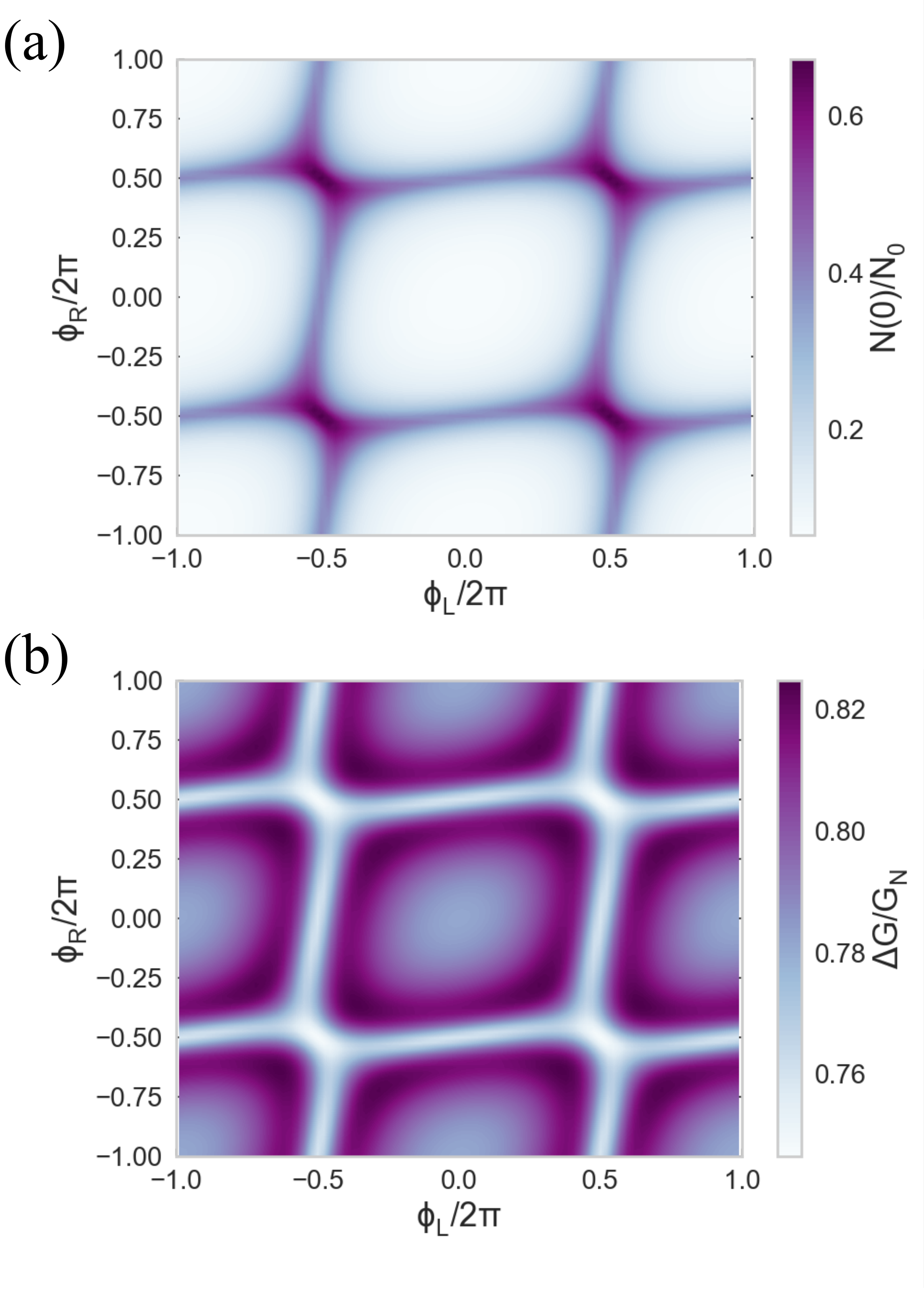}
\caption{(a) Calculated normalized density of states at the Fermi energy $N(0)$ as a function of the phase differences $\phi_L$ and $\phi_R$ for the device of Fig. \ref{fig:fig1}(a).  (b) Corresponding normalized change in conductance $\Delta G$ between the two normal reservoirs. The density of states plotted corresponds to the midpoint of the sample.}
\label{fig:fig3}
\end{figure}

Figure \ref{fig:fig3}(a) shows the density of states at the Fermi energy $N(0)$ at the midpoint of the path between the two normal reservoirs as a function of $\phi_L$ and $\phi_R$ for the device geometry of Fig. \ref{fig:fig1}(a).  Figure \ref{fig:fig3}(b) shows the corresponding variation in charge conductance $\Delta G$ divided by the normal state conductance $G_N$.  The conductance is calculated by applying a small voltage bias between the two normal contacts and determining the current flowing in the normal wire.  Experimentally, one might impose the phases $\phi_L$ and $\phi_R$ by connecting the appropriate superconducting contacts with loops and threading magnetic fluxes through them.  This means that the sum of the supercurrents entering the superconducting contacts at $\phi_L$ and $\phi_R$ must equal the supercurrent exiting the superconducting contact at $\phi =0$, a condition we impose explicitly in performing the calculation.  

Note that at this location in the device of Fig. \ref{fig:fig1}(a), the gap never closes, i.e., there is a gap in $N(0)$ for all values of $\{\phi_L,\phi_R\}$, with the minimum gap arising at $\{\phi_L,\phi_R\}=\{\pi, \pi\}$.  If we consider $N(0)$ at some other location, for example at a point in the normal wire midway between two superconducting reservoirs, one might observe a larger variation of $N(0)$ with $\{\phi_i\}$ (see Figs. \ref{fig:fig2}(a) and (b)).  The conductance, which is sensitive to the entire sample, shows a substantial enhancement but the relatively small variations ($\sim 10$\%) characteristic of the proximity effect over the entire phase space, as seen in Fig. \ref{fig:fig3}(b).  Nevertheless, comparison of Figs. \ref{fig:fig3}(a) and (b) shows that there is a strong correlation between $\Delta G$ and $N(0)$.  $\Delta G$ does have its largest value where $N(0)$ has the smallest gap overall, which gives hope that conductance measurements can be used to map full gap closures for suitably designed device geometries.      

\begin{figure}
\includegraphics[width=8cm]{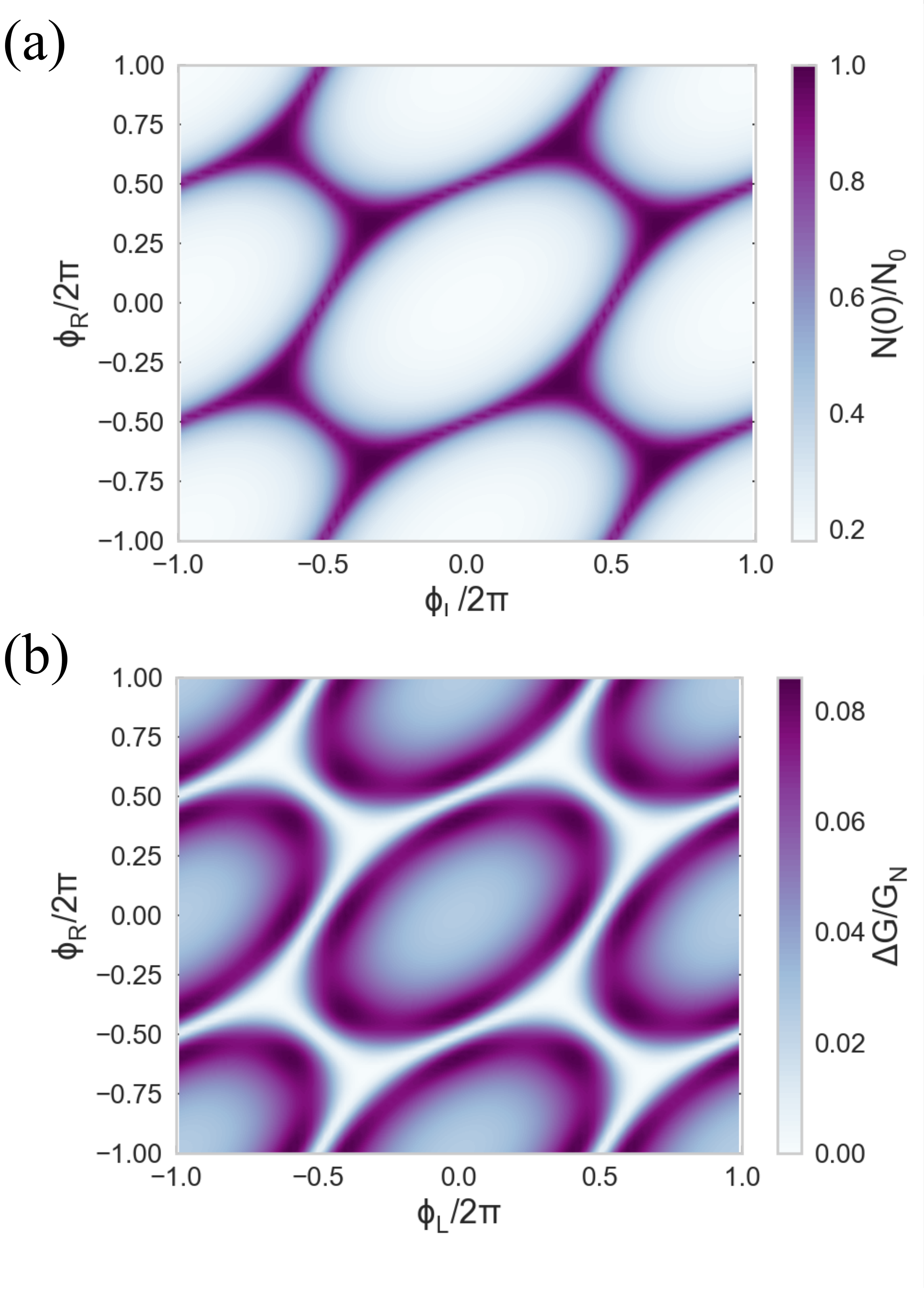}
\caption{(a) Calculated normalized density of states at the Fermi energy $N(0)$ as a function of the phase differences $\phi_L$ and $\phi_R$ for the device of Fig. \ref{fig:fig1}(b).  (b) Corresponding normalized change in conductance $\Delta G$ between the two normal reservoirs.  The density of states plotted corresponds to the midpoint of the sample.}
\label{fig:fig4}
\end{figure}

One such geometry is represented by the device of Fig. \ref{fig:fig1}(b), for which the corresponding $N(0)$ and $\Delta G$ maps are shown in Figs. \ref{fig:fig4}(a) and (b) respectively.  The first thing one notices is that these plots are quite clearly very different from those in Fig. \ref{fig:fig3}, both in their symmetry and in their magnitude variations.  $N(0)$ varies over a wide range, and most importantly, the gap in the density of states completely closes for certain points in $\{\phi_L,\phi_R\}$.  Surprisingly, these are not at odd half-multiples of $\phi/2 \pi$ (e.g., $\{\phi_L,\phi_R\}=\{\pi, \pi\}$) but instead occur at pairs of points close to these values, which explains why the spatial map of $N(E)/N_0$ taken at $\{\phi_L,\phi_R\}=\{\pi, \pi\}$ shown in Fig. \ref{fig:fig2}(d) is not completely flat.  Comparison to the corresponding conductance map (Fig. \ref{fig:fig4}(b)) shows that the conductance reaches its normal state value at these points.  The maximum conductance changes are still of the order of 10\%, but the key point is that at certain regions of flux space, the gap in the density of states completely closes, and that these regions can be identified by the fact that the conductance of the device becomes its normal state value.  

We note that the density of states plot shown in Fig. \ref{fig:fig4}(a) is very similar to the plots calculated by Amundsen \textit{et al.}\cite{amundsen} and Vischi \textit{et al.}\citep{vischi} for diffusive structures with 3 superconducting contacts, except that their device geometries did not have the additional normal reservoirs of our device geometry.  The common necessary ingredient required to obtain this density of states appears to be the three superconducting reservoirs connected to a single point of the device by normal wires of equal length.  This suggests that that a similar design be used for devices with a larger number of superconducting contacts, i.e., all the superconducting contacts be connected by normal wires to a single point on the device.

In summary, we have shown that measurements of the conductance of suitably designed devices can be used to explore the artificial band diagrams of diffusive multiterminal JJ structures, with the closing of the gap in the density of states manifesting itself through the conductance achieving its normal state value.  The simplicity of conductance measurements means that one can explore the evolution of the gap as the system traverses various trajectories in the corresponding quasimomentum phase space.

\appendix
\section{Details of numerical solutions}
The quasiclassical equations of superconductivity are solved in the so-called Riccati parametrization, in which the retarded Green's function of the system is represented in terms of the Riccati parameters $\gamma$ and $\tilde{\gamma}$ as
 \begin{equation}
\qr_s = \frac{1}{1 + \gamma \tilde{\gamma}}   
\begin{pmatrix}
1-\gamma \tilde{\gamma} & 2 \gamma \\
2 \tilde{\gamma} & \gamma \tilde{\gamma} -1 )
\end{pmatrix}.
\end{equation}
From the Usadel equation we obtain 2 differential equations for $\gamma$ and $\tilde{\gamma}$ in the normal wires
\begin{subequations}\label{eqn:eqns2} 
\begin{align}
D \left[\der^2 \gamma - \frac{2 \tilde{\gamma}}{1 + \gamma \tilde{\gamma}} (\der \gamma)^2 \right] +  2 i E \gamma &= 0, 
\\
\intertext{and}
D \left[\der^2 \tilde{\gamma} - \frac{2 \gamma}{1 + \gamma \tilde{\gamma}} (\der \tilde{\gamma})^2 \right] +  2 i E \tilde{\gamma} &= 0. 
\end{align}
\end{subequations}

In terms of the Riccati parameters, the spectral supercurrent is given by
\begin{equation}
Q = 2 \Re \left[ \frac{1}{(1+ \gamma \tilde{\gamma})^2} (\gamma \der \tilde{\gamma} - \tilde{\gamma} \der \gamma) \right]
\end{equation}
where $\Re$ stands for the real part.  The relevant modified diffusion coefficents $M_{ij}$ are given by
\begin{subequations} 
\begin{align}
M_{33} & = \frac{1}{|1 + \gamma \tilde{\gamma}|^2} \left[(|\gamma |^2 + 1)(|\tilde{\gamma}|^2 +1) \right], \\
M_{03} & = \frac{1}{|1 + \gamma \tilde{\gamma}|^2} \left[|\tilde{\gamma}|^2 -|\gamma |^2 \right].
\end{align}
\end{subequations}
In terms of these parameters, the charge current is given by
\begin{equation}
j(R,T) = e N_0 D \int dE [M_{33}(\partial_R h_T ) + Qh_L + M_{03}(\partial_R h_L )]
\end{equation}
here $h_T$ and $h_L$ are the transverse and longitudinal 	quasiparticle distribution function.  Finally, the density of states is given by
\begin{equation}
N(E) = \frac{1 - |\gamma|^2 |\tilde{\gamma}|^2}{|1 + \gamma \tilde{\gamma}|^2}.
\end{equation}  

The Usadel equations and corresponding kinetic equations are solved subject to certain boundary conditions. Details of these boundary conditions as well as the solution procedure can be found in the appendix of Ref. [\citenum{noh}].

\end{document}